\documentclass{aa}  
\usepackage{color}
\usepackage{soul}
\usepackage{graphicx}
\usepackage[
    colorlinks=true,
    pdfborder={0 0 0},
    linkcolor=blue
]{hyperref}

\usepackage{txfonts}
\usepackage{natbib}
\bibpunct{(}{)}{;}{a}{}{,} 

\begin{document}

   \title{Toward a large bandwidth photonic correlator for infrared heterodyne interferometry}
   \subtitle{A first laboratory proof of concept}
   \author{G. Bourdarot\inst{1,2}, 
          H. Guillet de Chatellus\inst{2} \and
          J-P. Berger\inst{1}
          }
    \titlerunning{A large bandwidth photonic correlator for infrared heterodyne interferometry}
    \authorrunning{Bourdarot, Guillet de Chatellus and Berger}

   \institute{
   Univ. Grenoble Alpes, CNRS, IPAG, 38000 Grenoble, France
    \and
    Univ. Grenoble Alpes, CNRS, LIPHY, 38000 Grenoble, France
             }

   \date{Received 20 December 2019 / Accepted 20 May 2020}

 
  \abstract
   {Infrared heterodyne interferometry has been proposed as a practical alternative for recombining a large number of telescopes over kilometric baselines in the mid-infrared. However, the current limited correlation capacities impose strong restrictions on the sensitivity of this appealing technique.}
   {In this paper, we propose to address the problem of transport and correlation of wide-bandwidth signals over kilometric distances by introducing photonic processing in infrared heterodyne interferometry.}
   {We describe the architecture of a photonic double-sideband correlator for two telescopes, along with the experimental demonstration of this concept on a proof-of-principle test bed.}
   {We demonstrate the \textit{a posteriori} correlation of two infrared signals previously generated on a two-telescope simulator in a double-sideband photonic correlator. A degradation of the signal-to-noise ratio of $13\%$, equivalent to a noise factor $\text{NF}=1.15$, is obtained through the correlator, and the temporal coherence properties of our input signals are retrieved from these measurements. }
   {Our results demonstrate that photonic processing can be used to correlate heterodyne signals with a potentially large increase of detection bandwidth. These developments open the way to photonic processing of wide bandwidth signals for mid-infrared heterodyne interferometry, in particular for a large number of telescopes and for direct imager recombiners.}


   \maketitle
%

\section{Introduction}

Optical interferometry and Very Long Baseline Interferometry (VLBI) are the two techniques that currently achieve the highest angular resolution in astronomy. The scale-up of infrared interferometry to an imaging facility with milli-arcsecond resolution and below represents a long-term objective of major interest for astrophysics \citep{monnier_planet_2018}. Such an instrument would require a large number of telescopes (N $\geq$ 12), in order to obtain a (u,v)-coverage that is compatible with imaging, and a kilometric baseline, in order to reach milli-arcsecond resolution in the near- and mid-infrared. At the present time, current facilities have the capacity to recombine up to four telescopes in the near- and mid-infrared at the Very Large Telescope Interferometer (VLTI) \citep{lopez_overview_2014} and up to six telescopes in the near-infrared at the CHARA array \citep{che_imaging_2012}; these two facilities have a maximum baseline of 130 m and 330 m, respectively. However, the extension of this current direct detection scheme represents a major technical challenge, in particular because of the infrastructure requested to operate the vacuum delay lines and the recombination of a large number of telescopes, which cannot necessarily be extrapolated from current existing infrastructures. \\ 
In this context, heterodyne detection, in which incident light is coherently detected on each telescope, has been proposed as a potential alternative in the mid-infrared \citep{townes_spatial_1984,swenson_radio-astronomy_1986,ireland_dispersed_2014}. Although heterodyne detection is commonly used in the radio to submillimeter domain, its extrapolation to higher frequencies (1 THz to several 10s THz) is limited by a radically different instrumentation compared to radio and submillimetric interferometry, and more fundamentally, by its lack of sensitivity at higher frequency. There are two reasons for this lack of performance. First, at equal bandwidth, there is a relative penalty in signal-to-noise ratio (S/N) between direct and heterodyne detection due to the fundamental quantum noise in heterodyne detection, which is a degradation that has been estimated to be on the order of $\sim 40$ by \citet{hale_berkeley_2000}. Second there is a very narrow instantaneous detection bandwidth in heterodyne detection (a few GHz typically) compared to the frequency (30 THz at $10\,\mu\text{m}$) of the incident radiation. On the other hand, heterodyne detection offers the advantage of recombining a large number of telescopes without a loss in S/N in contrast to direct detection.\\
The work presented in this paper should be placed in the context of a global effort to examine how present-day technology allows us to revisit the true performance of a mid-infrared heterodyne astronomical  interferometer composed of tens of telescopes and how it can be fairly compared with a direct interferometry approach. In this work, we do not attempt a full comparison, that we reserve to a forthcoming paper. We do explore one novel approach to one of the building blocks of such an interferometer : the correlator.\\
Following the idea laid out by \citet{swenson_optical_1986} and \citet{ireland_dispersed_2014} we propose that part of the sensitivity issue of  the heterodyne concept related to the bandwidth limitation can be overcome by using synchronized laser frequency combs as local oscillators (LOs) and detectors with much higher bandwidths. In this framework, the incoming celestial light interferes with a frequency comb and is dispersed to sample tens to hundreds of adjacent spectral windows. In addition, progress in mid-infrared technology has recently led to spectacular improvement of more than an order of magnitude of the detection bandwidth, in particular with the emergence of graphene detectors \citep{wang_room-temperature_2019}, which have  a  frequency  response of up  to  40  GHz,  and  quantum  well infrared photodetectors (QWIP) \citep{palaferri_room-temperature_2018} demonstrated  at  20  GHz. These developments bear the promise of even higher bandwidths of up to 100 GHz, more than an order of magnitude larger than what has been used on sky. As a consequence, as pointed out by \cite{ireland_dispersed_2014}, this detection scheme raises the formidable challenge of correlating thousands of pairs of signals.\\
The three-beam Infrared Spatial Interferometer (ISI) was based on the use of an analog radio frequency (RF) correlator with an input bandwidth ranging from 0.2 GHz to 2.8 GHz, using passive RF components. In the same way, Cosmological Microwave Background (CMB) interferometry has a long history of developing analog RF wideband correlators \citep{dickinson_cmb_2012}; an analog lag-correlator design recently reached up to 20 GHz bandwidth \citep{holler_2-20_2011}. Although these developments in CMB interferometry could constitute immediate attractive solutions, several difficulties inherent to wideband RF technology limit its use in the short and medium term, for infrared interferometry. The 20 GHz correlator presented in \cite{holler_2-20_2011} requires a specific RF design based on a custom-made Gilbert cell multiplier and Wilkinson splitter tree at the limit of the current technology and this design is unlikely to go far above 40 GHz any time soon. Parasitic frequency, although not a fundamental limit, could also turn out to be a disadvantage of wideband RF systems.
On the numerical side the currently most advanced digital correlation systems are those developed for the Northern Extedend Array (NOEMA) \citep{gentaz_polyfix_2019}  or for the Atacama Large Millimiter Array (ALMA) \citep{escoffier_alma_2007}. For NOEMA, the PolyFIX correlator currently accepts the widest instantaneous bandwidth per antenna. The PolyFIX correlator can process 32 GHz wide digitized signals coming from 12 antennas (8 GHz per receptor for the  two polarizations and two sidebands). ALMA correlator can process 8 GHz wide signals coming from up to 64 antennae.
Both approaches are worth exploring for infrared interferometry when considering an array of a few telescopes that have detector instantaneous bandwidths of a few  10 GHz and only a few spectral channels. However, their extrapolation to instantaneous bandwidths of 50 GHz to 100 GHz such as the bandwidths expected with new generation detectors, tens of telescopes and tens to hundreds of spectral bands call for a different approach. As evaluated in \citet{ireland_dispersed_2014}, this requires custom made developments with computing power at least two orders of magnitude greater than the existing correlators.\\
In order to tackle this conundrum, we propose a photonic solution to the problem of the correlation of broadband RF signals. We exploit the old idea of transmitting RF signals over optical waveguides to encode the intermediate frequency (IF) beating between the incoming signal and the local oscillator onto a coherent optical carrier, which could then be processed by the means of photonic operations. Remarkably, the past decade has seen an impressive development of microwave photonics, which aims precisely at generating, routing, and processing broadband RF signals, using standard photonic techniques  \citep{capmany_microwave_2007,nova_lavado_millimeter-wave_2013}. The ability to couple such analog processing with optical transport over fiber, compatible with standard telecom components, provides the building blocks of an analog correlator.
\\
In this paper we introduce the idea of a correlator for infrared heterodyne interferometry that makes use of photonic phase modulators to encode the RF beating signal onto a coherent carrier. Our scheme is based on commercially available electro-optic phase modulators and fiber-optics components. These can handle up to 50 GHz (off the shelf) and bear the promises of hundreds of GHz capability \citep{burla_500_2019}. In Sect. \ref{sec:principles-photco}, we present the principles of a simple correlation architecture with two telescopes, which reproduces the equivalent function of the initial ISI analog correlator. The experimental results of a proof of principle of this concept are presented in Sect. \ref{sec:PoC}, where two heterodyne beating signals have been generated experimentally, and correlated a posteriori on a photonic correlator several months later. The perspective and limits of this technique are discussed in Sect.\ref{sec:discussion} . Conclusions are drawn in Sect.\ref{sec:conclusion} . A theoretical sensitivity study, taking into account the instrumental parameters of a practical infrastructure, the gain in detection bandwidth introduced by a photonic correlator, the extrapolation in a multiplexed architecture, and its comparison with a direct detection scheme for a large number of telescope will be the object of a follow-up paper.

\begin{figure*}[t!]
   \centering
   \includegraphics[width=17cm]{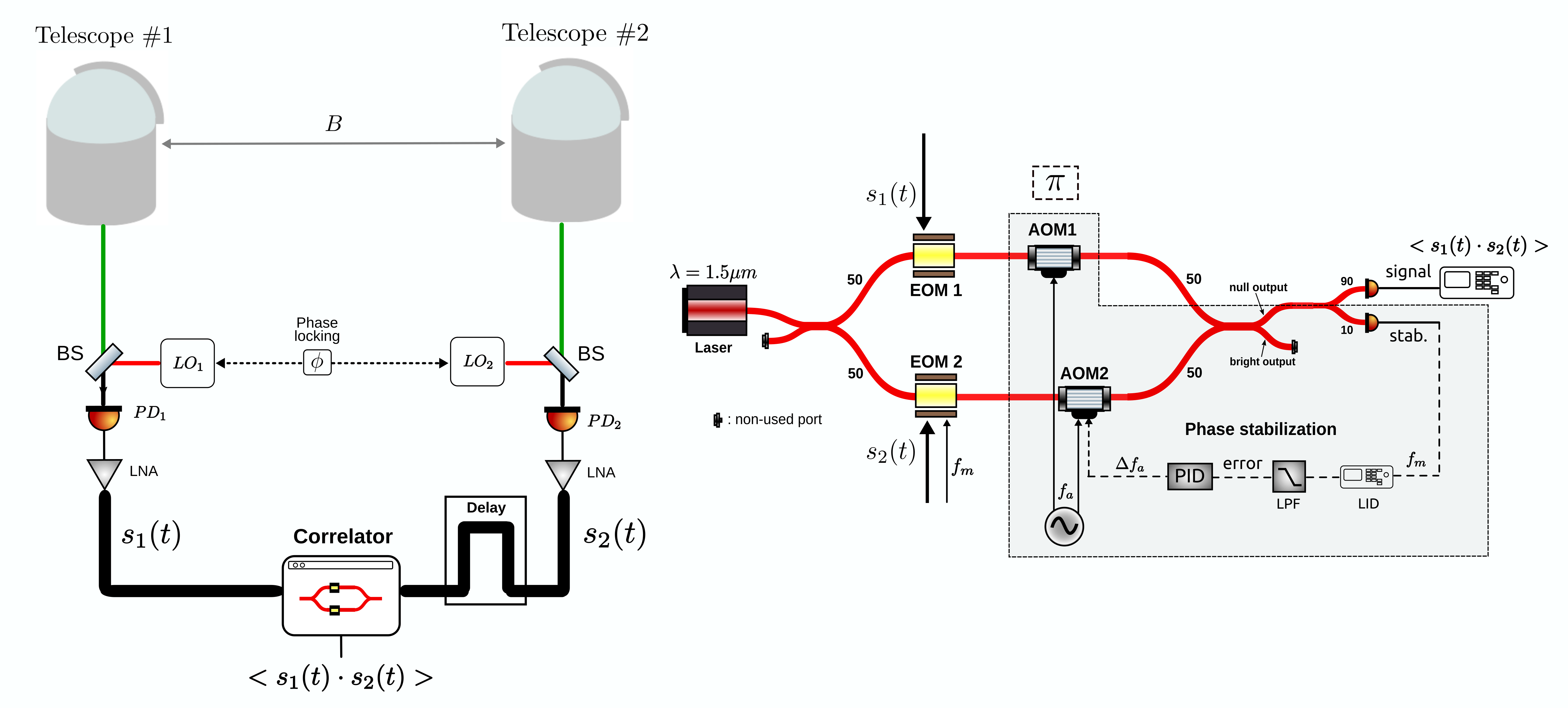}
      \caption{Left : General layout of a 2 telescopes heterodyne interferometer. Right: General layout of an optical correlator. A Mach-Zehnder interferferometer, stabilized at the zero null, is modulated by EOMs placed in each arm. In this configuration, the output of the photodiode contains a term proportional to the product of the RF signals at the input of each modulator. Abbreviations as follows: LO: local oscillator; BS: beam splitter; PD: photo-diode; LNA: low-noise amplifier; LID: lock-in detection; LPF: low-pass filter; PID: proportional integral derivator; AOM: acousto-optic modulator; EOM: electro-optic modulator.
              }
         \label{fig:principle}
\end{figure*}

\section{Principles of a photonic correlator}
\label{sec:principles-photco}
\subsection{Principles of infrared heterodyne interferometry}
As in radio astronomy, a single baseline heterodyne interferometer is composed of two distant telescopes on which the incident light is \textit{coherently} detected by its mixing with a stable frequency reference, referred to as the \textit{local oscillator} (LO), on a detector squaring the field. In the optical domain, the LO is a laser, and the detector a fast photodiode. Assuming a local oscillator $E_{LO}$ and an incident field $E_S$ the heterodyne signal at each telescope is written as

\begin{align}
i_k&\propto |E_S(t)+E_{LO}(t)|^2 \nonumber \\
&\propto| |E_S|e^{-i\left( (\omega_{LO}\pm\omega_{IF}) t+\phi_S(t)\right)}+|E_{LO}|e^{-i(\omega_{LO} t+\phi_{LO}(t))} |^2 \nonumber\\
&\propto|E_S|^2+|E_{LO}|^2+2|E_{S}||E_{LO}|\operatorname{Re}{\left(e^{-i( \pm\omega_{IF} t+(\phi_S(t)-\phi_{LO}(t))}\right),}
\label{eq:basic}
\end{align}{}

where $\omega_{LO}$ is the laser angular frequency, $\omega_{IF}$ the IF detected in the RF range, and $\phi_S$ and $\phi_{LO}$ the phases of the signal and LO, respectively. The signal angular frequency is denoted as $\omega_S=\omega_{LO}\pm\omega_{IF}$ to highlight the lower and upper sidebands of the signal; these are downconverted at the same IF $\omega_{IF}$.
The measured optical intensity thus contains a beating term proportional to the electric field, enabling the detection of the phase. After filtering through the detection chain, with transfer function $H(\omega)$, the beating term $s_k(t)$ is written \citep{boyd_radiometry_1983} as

\begin{equation}
s_{k}(t)=2H(\omega_{IF})\cdot|E_S||E_{LO}|\cos{\left(\pm\omega_{IF}t+(\phi_S(t)-\phi_{LO}(t))\,\right)}
.\end{equation}{}    

As these signals coming from each telescope are proportional to the input electric field, their multiplication is proportional to the coherent flux $Fe^{i\phi_o}$ of the source, where $\phi_o=2\pi\frac{\overrightarrow{B_p}\cdot\overrightarrow{\sigma}}{\lambda}$ the phase of the astrophysical object, $\overrightarrow{B_p}$ the projected baseline, $\overrightarrow{\sigma}$ the angular coordinate of the object from the phase center, and $\lambda$ the central wavelength. More specifically, in the case in which the two sidebands $\pm\omega_{IF}$ are not separated, the product of the two voltages from each telescope is \citep{thompson_interferometry_2017,monnier_infrared_1999} 
\begin{align}
r(t)&=<s_1(t)\cdot s_2(t)>\nonumber\\
&=C|F||G(\tau)|\cos(\phi_o+\Delta\omega_{LO}t)\cos(\phi_G+\omega_c\tau),
\label{eq:DSB}
\end{align}{}
assuming the detection bandwidth could be modeled with a rectangular filter function of width $\Delta\omega$, centered around $\omega_c$, where $\tau$ is the relative delay between the two optical signals, $\Delta\omega_{LO}$ the frequency difference between the LOs, $C$ a constant, and $|G(\tau)|e^{i\phi_G}=\frac{1}{2\pi}H^2_0\Delta\omega\left[\frac{\sin(\Delta\omega\tau/2)}{\Delta\omega\tau/2}\right]$ the frequency response of the detection chain that has an amplitude $H_0$. This expression corresponds to the signal of a double-sideband correlator (DSB), in which the fringes are modulated at the frequency $\Delta\omega_{LO}$. Importantly, we assume in Eq. (\ref{eq:DSB}) that the relative phase between the LOs $\Delta\phi_{LO}=\phi_{L01}(t)-\phi_{L02}(t)$ is null and stable over the time of detection, that is, that the LOs are phase-locked to each other. In practice, this phase-locking can be obtained either by distributing the same LO or by measuring a beating signal between each distant LOs, in both cases on a phase-stabilized link. In addition, in the following, the object phase $\phi_o$ is assumed to be constant, that is, the atmospheric piston fluctuations are assumed to be negligible during an integration time.

\subsection{Principles of a double-sideband photonic correlator}
\label{sec:principles-photonic}
In its simple form, the function required at the level of the correlator thus consists in multiplying two input signals with a very wide bandwidth. In this section, we show that this multiplication product can be achieved with a simple photonic design.\\ 
We consider a Mach-Zehnder interferometer, as represented in Fig.\ref{fig:principle}, in each arm of which is inserted a phase modulator with a characteristic voltage $V_{\pi}$. In a phase modulator, the $V_{\pi}$ is defined as the equivalent tension for which a phase shift of $\pi$ is introduced. Each phase modulator transposes the wide bandwidth RF signal coming from a telescope onto a monochromatic optical carrier. Assuming that the voltage amplitude is small compared to $V_{\pi}$, and writing $\beta=\frac{\pi}{V_{\pi}}$, the optical field after each phase modulators is

\begin{equation}
E_k(t)=E_0 e^{i(\omega_0 t+\phi_k+\beta s_k(t))}\approx E_0 e^{i(\omega_0 t+\phi_k)}(1+i\beta s_k(t))
\label{eq:phase-mod}
.\end{equation}{}
If a total relative phase shift of $\Delta\phi=\phi_2-\phi_1=\pi$ is applied between the arms, the interferometer is placed in a quadratic regime and the output intensity of the Mach-Zehnder can be simply written as 
\begin{align}
i(t)&=|E_1(t)+E_2(t)|^2\nonumber\\
&=|E_0|^2\beta^2\Big(s_2(t)-s_1(t)\,\Big)^2\nonumber\\
&=|E_0|^2\beta^2\left(s^2_1(t)+s^2_2(t)-2s_1(t)s_2(t)\right).
\label{eq:beating}
\end{align}{}

We note that if, for example, a phase shift of $\pi$/2 was used, there would not be a beat signal $s_1\cdot s_2$ between the two signals in the output. The two first quadratic terms appear as noise signals spread out over the wide frequency range of phase modulators. In turn, the last term is the product of the incident signal coming from the telescope, which is proportional to the coherent flux, as described in Eq. (\ref{eq:DSB}). In the case in which $\Delta\omega_{LO}\neq0$, the DSB product signal is modulated at the frequency $\Delta\omega_{LO}$, and thus gives access to a measurement of the coherent flux of the interferometer. This fringe peak can be integrated over a very restricted frequency range around $\Delta\omega_{LO}$, in which the relative contribution of the quadratic terms $s^2_k(t)$ can be neglected. In the above developments, it is fundamental to note that the total bandwidth is now limited by the bandwidth of the phase modulators. In practice, current standard off-the-shelf, fibered, electro-optic modulators (EOMs) at telecom wavelength reach a bandwidth of 50 GHz, and EOMs with flat-frequency response beyond 500 GHz have been demonstrated \citep{burla_500_2019}. Such bandwidths would represent a crucial improvement of the input bandwidth at the level of the correlator.

\subsection{Signal distribution and phase stabilization} 
In this correlation scheme, the telescope signals are converted at the level of each telescope on an optical carrier by means of an EOM. The signal could then propagate through telecom fibers over kilometric distance, avoiding the problem of bandwidth limitations. This scheme is only possible under the condition in which the optical link is phase-stabilized over large distance to guarantee a stable functioning point at the Mach-Zehnder's null; this stresses the importance of a robust phase stabilization scheme.\\  
Given the similarities of the photonic correlation with the principle of operation of a nulling interferometer, the phase stabilization scheme developed in this frame \citep{gabor_stabilising_2008} could be adapted in the present case. However, stabilization through phase modulation, by the use of EOMs or a fiber stretcher, could only be applied on a limited optical path difference (OPD) range, which may be a limit for kilometric optical links. Alternatively, in Sect \ref{sec:phase-stab}, we detail the principle of a fast phase stabilization scheme of the null based on frequency modulation of the optical carrier, which can correct an arbitrary OPD amplitude variation.

\section{Proof of concept and practical implementation}
\label{sec:PoC}
In this section, we present the proof of principle of a DSB photonic correlator dedicated to infrared heterodyne interferometry. In Sect.\ref{subsec:2T}, we detail the test bed used on a broadband laboratory source to generate an equivalent heterodyne signal of a two element interferometer. Sect.\ref{sec:exp-impl} describes the practical implementation of the photonic correlator and its phase stabilization scheme based on frequency modulation. In Sect.\ref{sec:results}, we finally present the \textit{a posteriori} correlation through this photonic correlator of the two signals previously generated on the two-telescope test bed, together with a measurement of the temporal coherence of the broadband source initially used, and we provide an estimation of the S/N degradation through the correlator.

\subsection{Two telescope heterodyne signal generation}
\label{subsec:2T}
The general purpose of the two-telescope simulator is to produce a correlated signal on two seperated detectors, whose signal could reproduce the beating between a broadband source of radiation and two LOs with a stable relative phase. The experiment was carried out at telecom wavelengths for practical reasons, but could be generalized to other optical wavelengths, in particular the N band, which is the target of the present study. We emphasize that the purpose of this test bed was not to evaluate the sensitivity limit of a complete detection chain from the detectors to the output of the photonic correlator, which would necessitate dedicated mid-infrared detectors and LOs, but to produce representative correlation signals in terms of coherence properties at the entrance of the photonic correlator. We acknowledge that typical astrophysical sources in the near- (H band) or mid-infrared (N band) are significantly fainter than in this proof of concept. The characterization of a complete mid-infrared detection chain, on objects at the detection limit and low S/N, would be the next step of this study.

\begin{figure}[h!]
   \centering
   \includegraphics[width=9cm]{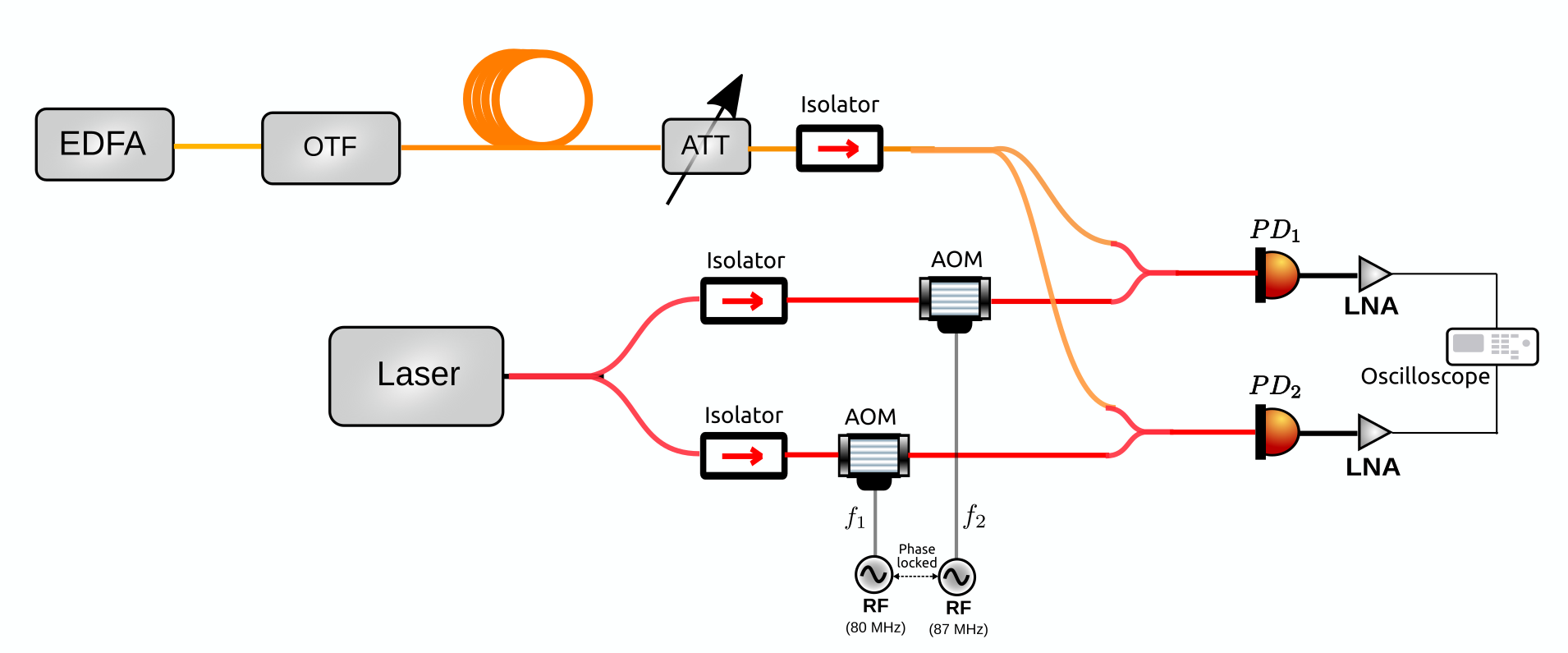}
      \caption{Scheme of the simulator of a 2 telescopes interferometer at telecom wavelength. A shift of $\Delta f=7\,\rm{MHz}$ is introduced between each LO by the use of AOMs driven at $f_1=80\,\rm{MHz}$ and $f_2=87\,\rm{MHz}$. The signal registered at the output of each PD has then been regenerated and correlated a posteriori on the optical correlator.
              }
         \label{fig:2T-testbed}
   \end{figure}
   
This test bed is described on Fig. \ref{fig:2T-testbed}. 
The representative elements of a two-telescope interferometer in the test bed are the following : \\
\indent \textit{Local oscillator:} A laser at $1.55\,\mu\textrm{m}$ is separated in two arms. Given the sub-kHz linewidth of the laser, the two equivalent LOs distributed on each arm are naturally in phase at the timescale of the measurement. In addition, as in ISI, a small frequency difference is applied between each arm by means of two acousto-optic modulators (AOMs). As this fequency difference is also the modulation frequency of the fringes, it has been experimentally set to $\Delta f=f_2-f_1=7\,\text{MHz}$, which is a spectrum region in which parasitic RF frequencies were absent. Since these modulators are designed so as to operate at $80 \pm 10\, \text{MHz}$, their frequency are set to $f_1=80\,\text{MHz}$ and $f_2=87\,\text{MHz}$. \\
\indent \textit{Broadband source:} We used an erbium-doped fiber amplifier (EDFA), without an input signal, as a broadband input source. An EDFA is a pumped gain medium, usually used in telecom to amplify an incident radiation. Without any input, it emits a broadband light spectrum through amplified spontaneous emission (ASE) of radiation. The ASE then passes through an optical tunable filter (OTF) adjusted to the few GHz bandwidth of the detector to limit the shot noise associated with the incident source. This source is finally divided in two arms, and distributed to two detectors. Once again, we emphasize that this source of radiation was not used to evaluate the sensitivity limit of a heterodyne detection in the near-infrared, but to reproduce representative coherence properties of a heterodyne signal.\\
\indent \textit{Detection:} The local oscillator and the input broadband source signal are combined and detected on two separate fast detectors. As a first step, a correlation peak at $7\,\text{MHz}$ was directly observed with an RF mixer, which multiplies the output of the two detectors. Multiple RF cables were successively used to introduce a delay $\sim1/\Delta\nu$ to scan the coherence length and to assess that the signal was not a parasitic frequency of the setup. In a second step, the output of each detectors were simultaneously recorded on a fast oscilloscope at a sampling rate of $2\,\text{Gb/s}$, with an analog bandwidth $\Delta\nu=400\,\text{MHz}$, which is the upper bandwidth limit in our detection scheme. \\
\indent \textit{A posteriori generation:} Once registered, these two RF traces were electronically generated \textit{a posteriori} by arbitrary-waveform generators (AWGs) to perform the \textit{a posteriori} correlation on the photonic correlator. Given the limited memory of the AWG, a set of $2^{16}$ points were generated at a sampling rate of $50\,\text{MHz}$. Taking into account the dilatation factor between registration and regeneration, the peak frequency was thus placed at a frequency $175\text{kHz}$ after regeneration.

\subsection{Experimental implementation of the photonic correlator and phase stabilization}
\label{sec:exp-impl}
In this subsection, we detail the experimental implementation of the photonic correlator described in Section \ref{sec:principles-photco}. As this photonic processing is independent of the carrier wavelength, its implementation could greatly benefit from the development of fibered components from telecom standards, which were also used in this proof of principle. \\

\subsubsection{Photonic processing}
\label{sec:phot-proc}
The actual implementation is represented in Fig. \ref{fig:principle}. A sub-kHz linewidth laser at $1.55\,\mu\textrm{m}$ is equally divided into two arms with a 50:50 fibered splitter. Each arm is then modulated by an EOM, on which is applied the RF signal generated \textit{a posteriori} from one of the two-telescope simulator traces, as described in Sect. \ref{subsec:2T}. A feedback loop is used to stabilize the phase of the Mach-Zehnder, and the two arms are recombined with another 50:50 fibered coupler. Finally, at the null output of this fibered Mach-Zehnder, the flux is split in two parts with a 90:10 fibered splitter, where $90\%$ of the flux is sent to the signal photodiode and $10\%$ of the flux to a detector used in the stabilization loop detector. After the signal photodiode, the fringes are modulated at the frequency $f=175\,\text{kHz}$, which can either be registered on an ADC, a lock-in amplifier, or a Fourier-transform oscilloscope. We adopted the latter solution. \\

\begin{figure}[t!]
   \centering
   \includegraphics[width=9cm]{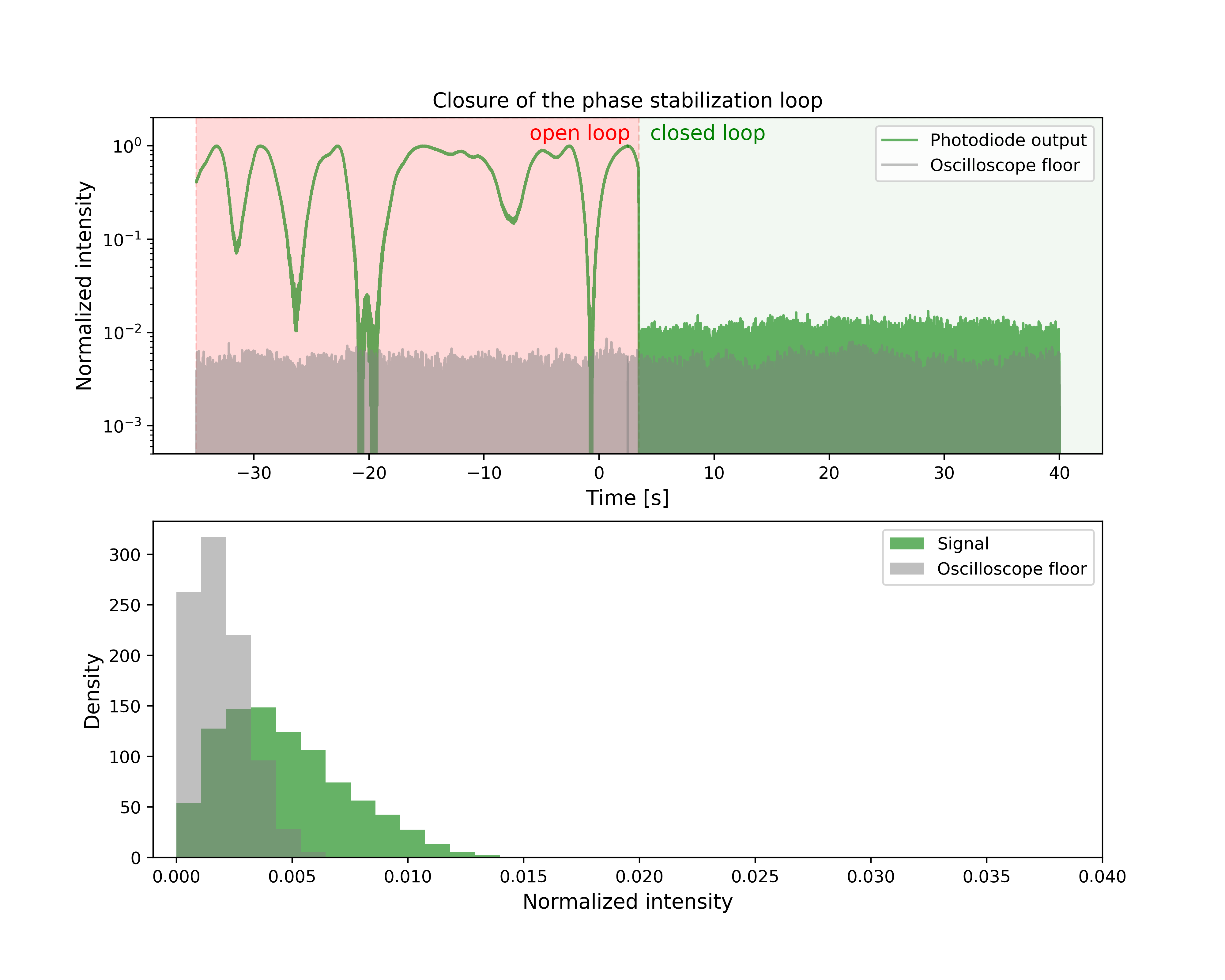}
      \caption{Phase stabilization scheme. Top: Open-loop is highlighted in  red, closed-loop in green. The photodiode output signal (green line) is just above the oscilloscope dark current (gray), which limits the effective contrast of the null. Bottom: Histogram of the photodiode output is shown. Asymmetric shape of the photodiode output is typical of a null output. 
              }
         \label{fig:phase-stab}
\end{figure}

\subsubsection{Phase stabilization loop}
\label{sec:phase-stab}
The general goal of the phase stabilization loop is to maintain the photonic correlator at the null intensity of the Mach-Zehnder. In the intensity null, the output intensity varies quadratically with the input voltage. The basic idea of the stabilization consists in generating a small amplitude phase modulation signal at a defined frequency $f_m$ in one arm, and using the real part of the first harmonic signal as an error signal to be minimized. Usually, the command signal is applied on a phase modulator (e.g., PZT, fiber stretcher, or EOM) to compensate for OPD variation. In this case, we set up a frequency modulation system, composed of two AOMs, where one is modulated in frequency by a proportional-derivative (PD) controller. Integrated over a small of time $\rm{d}t$, this frequency modulation $\Delta f_m$ acts as a phase modulation $\rm{d}\Phi_m=\Delta f_m \rm{d}t$, which is restricted neither in amplitude nor in speed in contrast to an OPD modulator system. Fig. \ref{fig:phase-stab} represents the closure of the phase stabilization loop. We estimated its stability to a mean phase deviation of $\overline{\phi}=\lambda /240$, and RMS deviation of $\sigma_{\phi}=\lambda/440$.\\
It has though to be noted that in such a frequency modulation scheme, large OPD drifts, on the order of a fraction of the coherence length $l_c=c/\Delta\nu$, have to be corrected by a dedicated OPD offset. However, given a maximum spectral bandwidth of $\Delta\nu\approx 100\,\text{GHz}$, the coherence length is on the order of millimeter scale, thus requiring only occasional offset correction after minute or hour timescales. \\

\subsection{Noise factor and temporal coherence}
\label{sec:results}

\begin{figure*}[t!]
   \hspace*{-0.85cm}
   \centering
   \includegraphics[width=20.5cm]{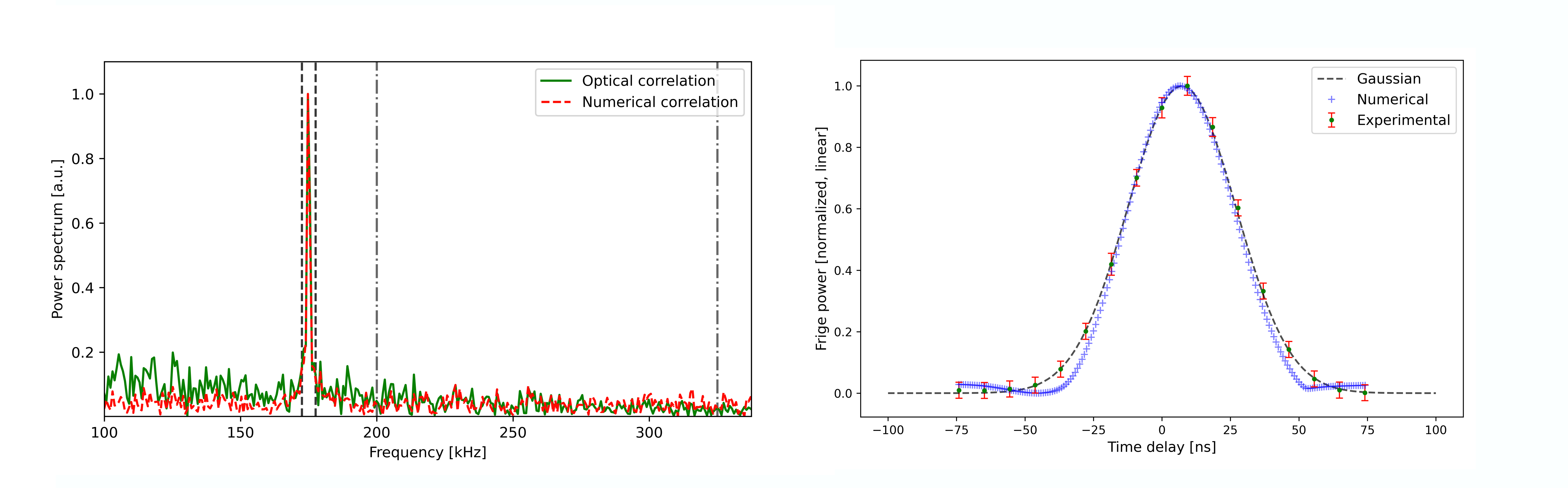}
      \caption{Left: Fringe peak at the ouptut of the correlator (y-axis, in linear scale). The red dashed line indicates the numerical correlation of the input signal. The green solid line shows the correlation of the signal through the optical correlator. Right: The coherence envelope of the fringe signal (green dot), and the envelope computed numerically (blue cross) are shown.  
      The experimental envelope fits a Gaussian profile, which differs from the numerical computation only at the feet of the envelope. This could be imputed both to the experimental measurement uncertainties and to the estimation of the impulse response of the AWG.}
         \label{fig:fringes}
   \end{figure*}
   
Once stabilized on the null, the signals \textit{a posteriori} generated by the AWGs are applied to the phase modulators. According to Eq. (\ref{eq:beating}), a fringe peak is observed at the modulation frequency $\frac{f_G}{f_S}\Delta f_{OL}=175\,\text{kHz}$, where $f_S$ is the recording sampling frequency and $f_G$ is the generation sampling frequency. This fringe signal is easily visible in the power spectral density of the photocurrent (PSD), as shown in Fig. \ref{fig:fringes}. In order to estimate the degradation introduced by the photonic correlator on the signal, we measured the noise factor, defined as the ratio of input and output S/N, as follows:
\begin{equation}
\text{NF}=\frac{(S/N)_{in}}{(S/N)_{out}}
.\end{equation}{} 
We estimated input S/N from the two recorded waveforms, by computing numerically the interference term in Eq. (\ref{eq:beating}). Fringe power and noise are estimated on two defined frequency windows, as shown if Fig. \ref{fig:fringes}, by computing the integrated power in the peak and the standard deviation of the noise floor, respectively. Output S/N is then estimated with the same method on the PSD of the photodiode output, on the same exact frequency windows. This analysis provides a ratio of the output S/N on the input S/N : $1/\text{NF}=87\% \pm 5\%$ that is, a S/N degradation of $13\%$, also corresponding to $\text{NF}=1.15$. This result is limited by a non-negligible oscilloscope dark current, as seen in the histogram of Fig. \ref{fig:phase-stab}, and a strong contribution of a low-frequency $1/f$, as visible in Fig. \ref{fig:fringes}, which artificially degrade the S/N of the fringe peak, but are not fundamentally due to the optical correlator. \\
In addition, we assessed the temporal properties of our correlation signal to observe its coherence envelope and give an additional verification that the fringe peak observed could not be produced by a parasitic signal. To do so, a numerical delay is introduced at the level of one AWG, for each value for which we measured the fringe peak power. The coherence envelope is shown on Fig. \ref{fig:fringes}, and superposed to the coherence envelope computed numerically. The experimental profile fits a Gaussian with a full width at half maximum (FWHM) $\tau\approx 20\,\text{ns}$, which corresponds to an equivalent bandwidth $\Delta f =1/\tau \sim 50\,\text{MHz}$. This is consistent with the maximum bandwidth of our regenerated signal with a sampling frequency $f_S=50\,\text{MS/s}$. Moreover, this measurement removes the possibility that the fringe peak is a parasitic signal.

\section{Discussion}
\label{sec:discussion}
\subsection{Further developments}
In this section, we discuss the further developments to be led in the path towards a practical correlator, dedicated to an imaging facility with kilometric baselines. In a second section, we discuss at a more general level the remaining open challenges of infrared heterodyne interferometry.
\subsubsection{Photometric calibration and delay compensation}
    In the continuity of our demonstration of signal correlation, the next steps of the development consist in measuring the spatial coherence of a laboratory object in the mid-infrared. For this purpose, a detailed procedure of photometric calibration will have to be carried out in order to normalize the coherent flux measured on the source and to deduce an estimate of the visibility. Furthermore, in this work we did not address the problem of delay compensation and earth rotation. Earth rotation translates into a phase velocity that can be computed and compensated at the level of a local oscillator by a dedicated frequency shift, which is also called lobe rotation. In addition, we did not address a group delay, which has to be compensated for in order to track the maximum of the correlation envelope within a coherence length $l_c$. This delay can be covered using a combination of switchable fibered delay, compensating for the large delays, and a continuously adjustable fibered delay line, covering small delays and relaxing the minimum resolution of the switchable module. We note that the design complexity of such a movable delay line, at telecom wavelength and on a very narrow spectral band, would be considerably lower than the design complexity of a direct mid-infrared vacuum delay line. As in telecom networks, dispersion could be managed with the use of dispersion compensating fibers, over a bandwidth of 100 GHz in this case, but with propagation distances significantly smaller than that encountered in telecom, up to a few kilometers in this case. The speed of the movable delay line could be relaxed by a careful control of the frequency shift of lobe rotation. 
    \subsubsection{Measurements with $\textrm{N}\geq 3$}
    This measurement with two telescopes could then lead to a generalization of the method to more than two telescopes, and in particular to the measurement of closure phases with three telescopes, in a way that is analogous to the method performed on the ISI correlator \citep{hale_techniques_2003}. We note that from the perspective of achieving image reconstruction with a large number of telescopes (N $\geq$ 12), the encoding of the signal on an optical carrier also offers the possibility to recombine all the fibers into an homothetic pupil plane, as in a Fizeau configuration, which would enable us to use the array in a direct imager mode. The transposition of the technique presented in this paper, where fringes are modulated at a given frequency ($7\,\text{MHz}$), may be adapted by lowering the modulation frequency to a rate compatible with a 2D-matrix acquisition rate (typically smaller than kHz), although this method does not seem optimal. Instead, direct phase stabilization scheme, as experimentally demonstrated in \citet{blanchard_coherent_1999} may possibly allow for direct imager acquisitions without applying a frequency modulation. Signal-to-noise preservation has, however, not yet been considered for this scheme, nor has it been demonstrated in \citet{blanchard_coherent_1999}.
 
\subsection{Open challenges of infrared heterodyne interferometry}
    Although we addressed the problem of correlation and signal transportation by the introduction of a photonic correlator and photonic processing, several challenges remain open on the path to a practical mid-infrared heterodyne interferometer. \\
    The first problem we did not address in the heterodyne system is the synchronization of distant  LOs separated by kilometric distances. We recall that this requirement concerns at least the relative phase between the different local oscillator, which has to be constant during a coherent integration time. For this purpose, the beating of each LO with an LO that serves as a master and reference can be used to apply a correction to each LO through a dedicated phase-lock-loop (PLL), a strategy that was implemented in ISI with up to three telescopes \citep{hale_berkeley_2000,hale_techniques_2003}. In practice, such a stabilization scheme would imply that we propagate each mid-infrared local oscillator on kilometric distances in our case, which imposes strong constraints in a practical infrastructure. The possibility to stabilize in phase each mid-infrared LOs with the distribution of a reference phase signal through a fiber link, in a way analogous to \citet{chanteau_mid-infrared_2013} for example, would substantially simplify the infrastructure of a mid-infrared heterodyne interferometer. \\
    The second problem that we did not address concerns the limit in sensitivity imposed by atmospheric phase fluctuations, which severely restricts the maximum coherent integration time. Previous studies \citep{ireland_dispersed_2014,ireland_status_2016} already raised this limitation, and proposed an out-of-band cophasing based on a companion instrument in the H band. Although this auxiliary instrument would be a direct interferometer, which is apparently in contradiction with the heterodyne detection scheme proposed, its implementation in the H band would surely be much easier than in the mid-infrared with the use of fiber components. In the case in which this atmospheric cophasing were absent, the \textit{heterodyne} interferfometer would still be functional, but limited to bright objects.\\
    Concerning the improvement of sensitivity, the introduction of this paper was based on the observation that current detectors now enable us to reach several tens of gigahertz of bandwidth in the mid-infrared. These compelling demonstrations will need further development to consolidate these results, in particular regarding the exact characterization and optimization of their quantum efficiency. \\
    Finally, as proposed in \cite{swenson_radio-astronomy_1986}, revived in \cite{ireland_dispersed_2014}, and from a more prospective view, a promising but difficult method to further increase the sensitivity of an heterodyne interferometer would consist in multiplexing a large number of LOs, with the associated number of detectors, to potentially obtain a spectral coverage comparable to direct detection. This method supposes the generation of mid-infrared frequency combs that have sufficient power per teeth, which constitutes a present active field of research. We note that such a multiplexed architecture could be advantageously coupled to a photonic correlation. \\

\section{Conclusions}
\label{sec:conclusion}
    Within the context of infrared heterodyne interferometry, we have introduced the use of photonic correlation in order to overcome the bandwidth limitation of the correlators developed so far. We proposed the architecture of a DSB correlator for two telescopes based on the use of a fibered Mach-Zehnder at telecom wavelength, precisely stabilized at the null of intensity, and demonstrated the \textit{a posteriori} correlation of two signals previously generated on a dedicated two-telescope test bed in the near-infrared. For this purpose, we realized a dedicated phase stabilization loop based on frequency modulation. The final photonic processing chain exhibits a degradation of the S/N of $13\%$, corresponding to a noise factor $\text{NF}=1.15$. The coherence properties of the initial input signals were also retrieved by introducing an incremental temporal delay. The next step of this development will consist in measuring the spatial coherence of an object in the mid-infrared with two telescopes, and to generalize this architecture to more than two telescopes and to the detection of closure phases. More generally, this proof of principle opens the way to the photonic processing and transportation of wide bandwidth signals for infrared heterodyne interferometry, which could constitute a valuable advance in the perspective of kilometric baseline interferometry with a large number of telescopes. 

\begin{acknowledgements}
We thank the anonymous referee for his/her constructive remarks that improved our initial manuscript. This work was supported by the Action Spécifique Haute Résolution Angulaire (ASHRA) of CNRS/INSU, co-funded by CNES. We acknowledge the funding of Agence Nationale de la Recherche ("ANR FROST", ANR-14-CE32-0022). The authors thank Jean-Baptiste Le Bouquin for his very relevant advice during the optimization of the set-up. The authors also thank Etienne Le Coarer for the always stimulating discussions with him and for having initiated our interest in this topic. 
\end{acknowledgements}

\bibliographystyle{aa} 
\bibliography{PaperI} 


\end{document}